\documentclass[12pt]{iopart}

\usepackage{graphicx,float,multirow,hhline,fancyvrb,graphics,indentfirst,varioref,wrapfig,lettrine,subfigure,cite}

\begin{document}

\title[Khalid et al.]{Quantification of Flow Noise Produced by an Oscillating Foil}

\author{Muhammad Saif Ullah Khalid${}^{1, 2}$, Xiaoping Jiang${}^{1}$, Imran Akhtar${}^{2}$, Binxin Wu${}^{1}$}
\address{${}^{1}$Research Center for Fluid Machinery Engineering and Technology, Jiangsu University, Zhenjiang, 212013, Jiangsu, People's Republic of China.\\
${}^{2}$Department of Mechanical Engineering, College of Electrical \& Mechanical Engineering, National University of Sciences \& Technology, Rawalpindi, 46000, Pakistan.}
\eads{\mailto{mskhalid1984@gmail.com}}
\footnote{This file is produced using the IOPScience LaTex class.}

\vspace{10pt}
\begin{indented}
\item[]June 2018
\end{indented}

\begin{abstract}
Getting inspired from swimming natural species, a lot of research is
being carried out in the field of unmanned underwater vehicles.
During the last two decades, more emphasis on the associated
hydrodynamic mechanisms, structural dynamics, control techniques
and, its motion and path planning has been prominently witnessed in
the literature. Considering the importance of the involved acoustic
mechanisms, we focus on the quantification of flow noise produced by
an oscillating hydrofoil here employed as a kinematic model for fish
or its relevant appendages. In our current study, we perform
numerical simulations for flow over an oscillating hydrofoil for a
wide range of flow and kinematic parameters. Using the
Ffowcs-Williams and Hawkings (FW-H) method, we quantify the flow
noise produced by a fish during its swimming for a range of
kinematic and flow parameters including Reynolds number, reduced
frequency, and Strouhal number. We find that the distributions of
the sound pressure levels at the oscillating frequency and its first
even harmonic due to the pressure fluctuations in the fluid domain
are dipole-like patterns. The magnitudes of these sound pressure
levels depend on the Reynolds number and Strouhal number, whereas
the direction of their dipole-axes appears to be affected by the
reduced frequency only. Moreover, We also correlate this emission
of sound radiations with the hydrodynamic force coefficients.

\end{abstract}

%
%
%
%
%

\section{Introduction}
\label{sec:intro} Since the advent of unmanned swimming
robots, underlying hydrodynamic mechanisms for producing the
required forces for propulsion and other movements have attracted the attention of researchers from diverse backgrounds. These belong to fluid
mechanics, fluid-structure interaction, dynamics and control, and
now machine vision etc. Extensive research work in all these fields
have been reported in the literature \cite{Dong2010, Boschitsch2014, Geng2015a, Fish2016a, Khalid2016a, Salzar2018}. However, this work lacks dedicated efforts in quantification of flow noise produced by such vehicles. It is particularly relevant for their prospective application to perform missions silently in the hostile environments. This study is also important in exploring the mechanisms adopted by fishes to feel the presence of other objects around them. They may utilize these signals for various biological motives such as behavioral communication and navigation in a certain environment. It is also interesting to examine the dependence of directionality in sound emission on flow properties and
body-kinematics. Many biologists have presented their work on sounds
produced by fish for their social and behavioral motives
\cite{Radford2011}. Also, they focused on investigating the hearing mechanisms utilized by different fishes. There were only a few studies in literature devoted to the quantification of sound produced by oscillating bodies \cite{Sueur2005, Bae2008, Inada2009, Lu2014, Geng2018}. These pieces of work concentrated on the noise production in air by flying species. Sueur et al. \cite{Sueur2005} characterized the acoustic radiations from the flapping appendages of a blowfly through close-range recordings. They found out that the sound spectrum contained a series of frequencies. They also observed that the sound pressure levels at first harmonic showed a dipole like structure in the horizontal plane of the fly, whereas second harmonic came out with a monopole like radiation pattern. In their case, fundamental harmonic was more dominant in the upstream direction of the fly, and second harmonic showed more dominance in the sideways. Their recordings in the downstream direction of the
specie corresponded to a broadband noise that was termed as the
pseudo-sound. They remarked that the pseudo-sounds were produced
from the vortices in the wake of the flying object. Bae et al. \cite{Bae2008} performed numerical simulations for flow around a
two-dimensional ($\mbox{2D}$) model of another flying insect in hovering and forward flight conditions. They determined two sound production mechanisms with dipole tones due to the traversing wing-motion and vortex edge scattering during a tangential motion. They argued that the primary tone with the wing beat frequency is directional for hovering motion, and it showed monopole like character for forward flight. For hovering, they also associated the tonal frequencies with those of the fluid force
coefficients, but the question of directionality for forward flight
remained unclear. Similarly, Inada et al. \cite{Inada2009} also
carried out three-dimensional ($\mbox{3D}$) simulations to uncover
the mechanisms for sound production by the flapping wings of three
insects; hawkmoth, honeybee, and fruitfly. They concluded that
flapping sound was dependent on amplitude, frequencies, and the
observer's position. They also found the fundamental harmonic as the
dominant one. The experimental work done by Lu et al. \cite{Lu2014}
on the flapping sound of a robot using flapping wings uncovered that a highly elastic material could reduce the flapping noise to a great extent. A more dedicated effort in this direction was recently done by Geng et al. \cite{Geng2018}. They performed numerical simulations using a
hydrodynamic/acoustic splitting approach, where acoustic field is
modeled through linearized perturbed compressible equations. Their
simulations revealed the directionality of sound, and reduction of
flapping sound with an enhanced flexibility of wings. Additionally,
they argued that the fundamental harmonic in the sound spectra dominates
in the direction of the stroke plane, and in the downstream
direction for rigid and flexible wings, respectively. In their case,
second harmonic showed more strength sideways for rigid wing, and
towards left and upper sides for the flexible wings. They also
explored that the direction of the dipole axis was different for
rigid and flexible wings. The sound emission and its propagation was
found to be highly associated with the wing loading. Up to the
authors' knowledge, no concerted effort to investigate the noise
produced in water by fish-bodies has been seen in the literature. Also, the dependence of the sound pressure levels on various key parameters has not been explored. Still, there are unanswered questions regarding the sound production and propagation from the oscillating body-structures which are more specific to fish swimming. These factors include the effect of flow parameters such as Reynolds number ($\mbox{Re}$) and the wake-width resulted due to vortices shed from the trailing-edge of the bodies,
and the kinematic parameters like oscillation/excitation frequency ($f_E$) and amplitude etc. These aspects motivate our current research work. This study will assist in understanding the mechanisms adopted by fish to avoid predators, and to propose better designs for unmanned underwater robots to perform their missions silently, and without being noticed by the surrounding hostilities.

In this study, we focus on quantifying the flow noise as the sound
pressure levels measured in $\mbox{dB}$ that is produced by the oscillatory motion of fish bodies, or their relevant appendages during swimming. As the hydrodynamic flow characterization for oscillating
bodies depends on three similarity parameters; the Reynolds number
($\mbox{Re}=\frac{{\rho}{U_\infty}{c}}{\mu}$), the reduced frequency
($k=\frac{{f_E}{c}}{U_\infty}$), and the Strouhal number
($\mbox{St}=\frac{{f_E}{A}}{U_\infty}$) \cite{Saadat2017}, the results
may be considered as applicable to the other oscillatory systems
operating with the same range of these parameters. Here, $\rho$,
$\mu$, $c$ and $U_\infty$ show the fluid-density, dynamic viscosity of the fluid, chord-length, and free-stream fluid-velocity,
respectively. The amplitude $A$ is the maximum amplitude traversed
by the trailing-edge of the hydrofoil, and a measure of the
wake-width for reverse von Karman vortex street. We carry out numerical
simulation for flow over an oscillating hydrofoil structure, a
representative of fish-body or its fin, for $\mbox{Re}=5,000$,
$10,000$, and $20,000$. We vary the $\mbox{St}$ ranging from $0.10$
to $0.50$ that is a usual range of fish swimming for efficient
hydrodynamic performance \cite{Saadat2017}. We also set $k$ equal to
$1$, $2.5$, and $5$ which represent its low, moderate, and high
values, respectively.

\section{Numerical Methodology}
\label{sec:nummethd}

\subsection{Hydrofoil Kinematics}
We model a fish-body, or its single fin using a hydrofoil having circular leading-edge with $D/c=0.06$ \cite{Boschitsch2014}.
Governing mathematical relation for its kinematics is given below.

\begin{eqnarray}
\theta(t)&={\theta_\circ}\sin{2{\pi}{f}{t}}
\label{eq:pitch}
\end{eqnarray}


\noindent where $\theta(t)$ is the instantaneous angular position of
the hydrofoil, $\theta_\circ$ is its maximum excursion from its mean
position, and $t$ denotes the time in seconds. This hydrofoil conducts pitching motion about its leading-edge.

\subsection{Flow Solver}
\label{subsec:solver} To perform our numerical simulations, we use
ANSYS Fluent $16.1$ \cite{Ansys}, a commercial finite volume based
computational platform, that has gained a lot of popularity among
researchers for several fluid flow simulations \cite{Kinsey2008,
Ashraf2011, Kinsey2014}. Incompressible unsteady Reynolds-averaged
continuity and Navier-Stokes (URANS) equations in Cartesian
coordinates, given as \ref{eq:cont} and \ref{eq:NS},
respectively, in tensor form, are solved through the pressure-based
solver.

\begin{eqnarray}
\frac{\partial u_j}{\partial x_j}&=0
\label{eq:cont}
\end{eqnarray}


\begin{eqnarray}
\frac{\partial u_i}{\partial t} + \frac{\partial}{\partial
x_j}({u_i}{u_j})&=-\frac{1}{\rho}\frac{\partial p}{\partial
x_i}+\nu \frac{{\partial^2}u_i}{{\partial x_j}{\partial x_j}}
\label{eq:NS}
\end{eqnarray}


\noindent where $x_j$ denotes the Cartesian coordinates, and
$j=\{1,2\}$. $u$ shows the Cartesian velocity components, $\rho$ is
the water density here, $p$ shows the pressure, and $\nu$ indicates
the kinematic viscosity. Due to the absence of a pressure term in
the Continuity equation (see \ref{eq:cont}), semi-implicit method
for pressure linked equations (SIMPLE), semi-implicit method for
pressure linked equations-consistent (SIMPLEC), pressure implicit
with splitting operators (PISO), and Coupled algorithms
\cite{Ansys} are provided in the solver. Although PISO is
recommended for unsteady flows \cite{The2017}, it is usually
advantageous when a large time-step (${\Delta}t$) is adopted to
computationally march in time. Hence, to improve the computational
efficiency, we adopt SIMPLE algorithm for our cases.

We utilize the Green-Gauss cell based technique for computation of the
gradient terms, the second order scheme for the convective pressure
terms, and the higher order Quadratic Upstream Interpolation for
Convective Kinematics (QUICK) scheme for the diffusion terms in the
momentum equation (\ref{eq:NS}). The unsteady term is
approximated by the second order implicit scheme. We perform the
unsteady simulations through the sliding mesh technique that allows
physical pitching motion of the hydrofoil without disturbing the original
mesh.

\begin{figure}
\center
\subfigure[]{\label{fig:grida}\includegraphics[scale=0.4]{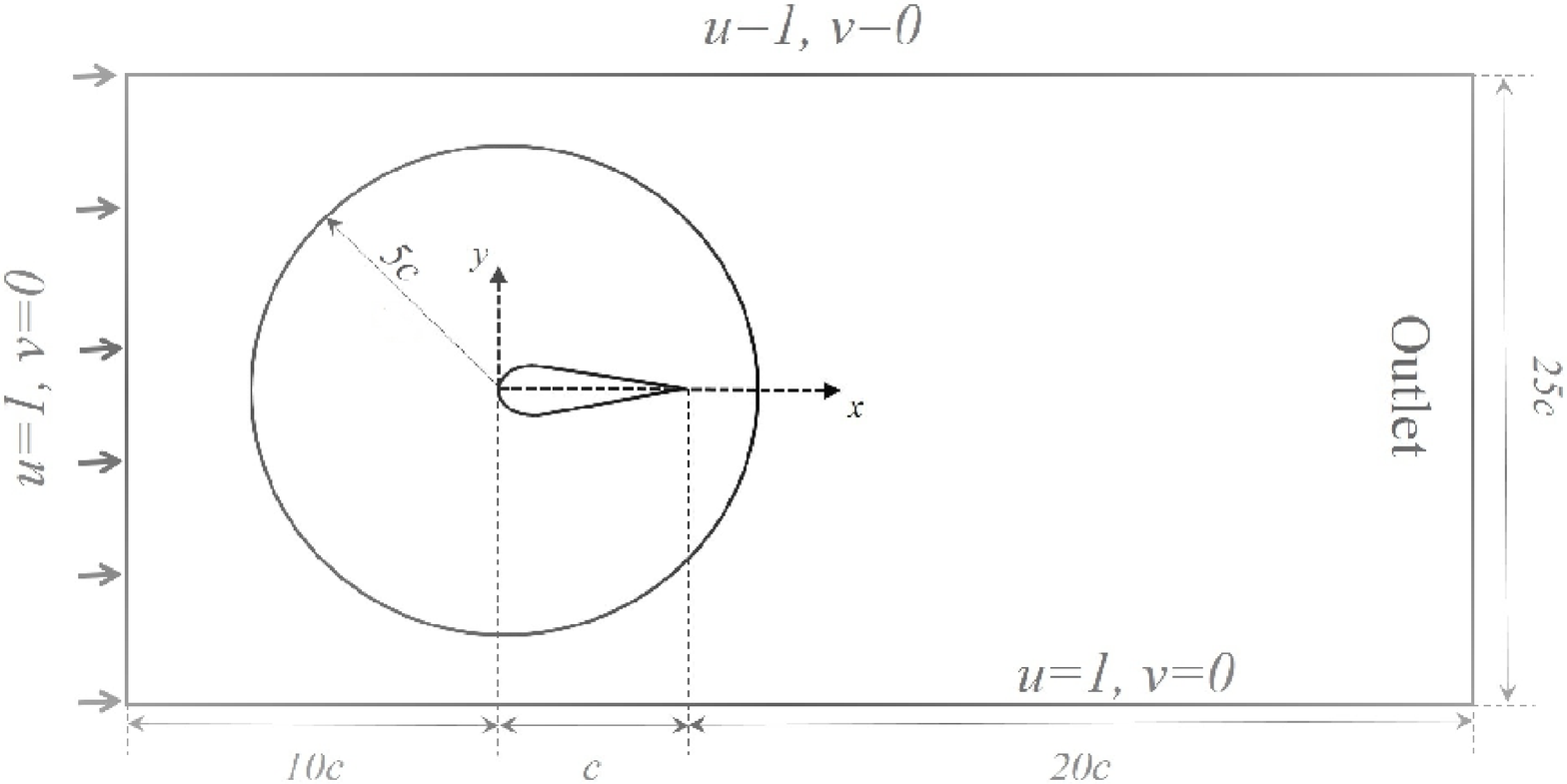}}
\subfigure[]{\label{fig:gridb}\includegraphics[scale=0.2]{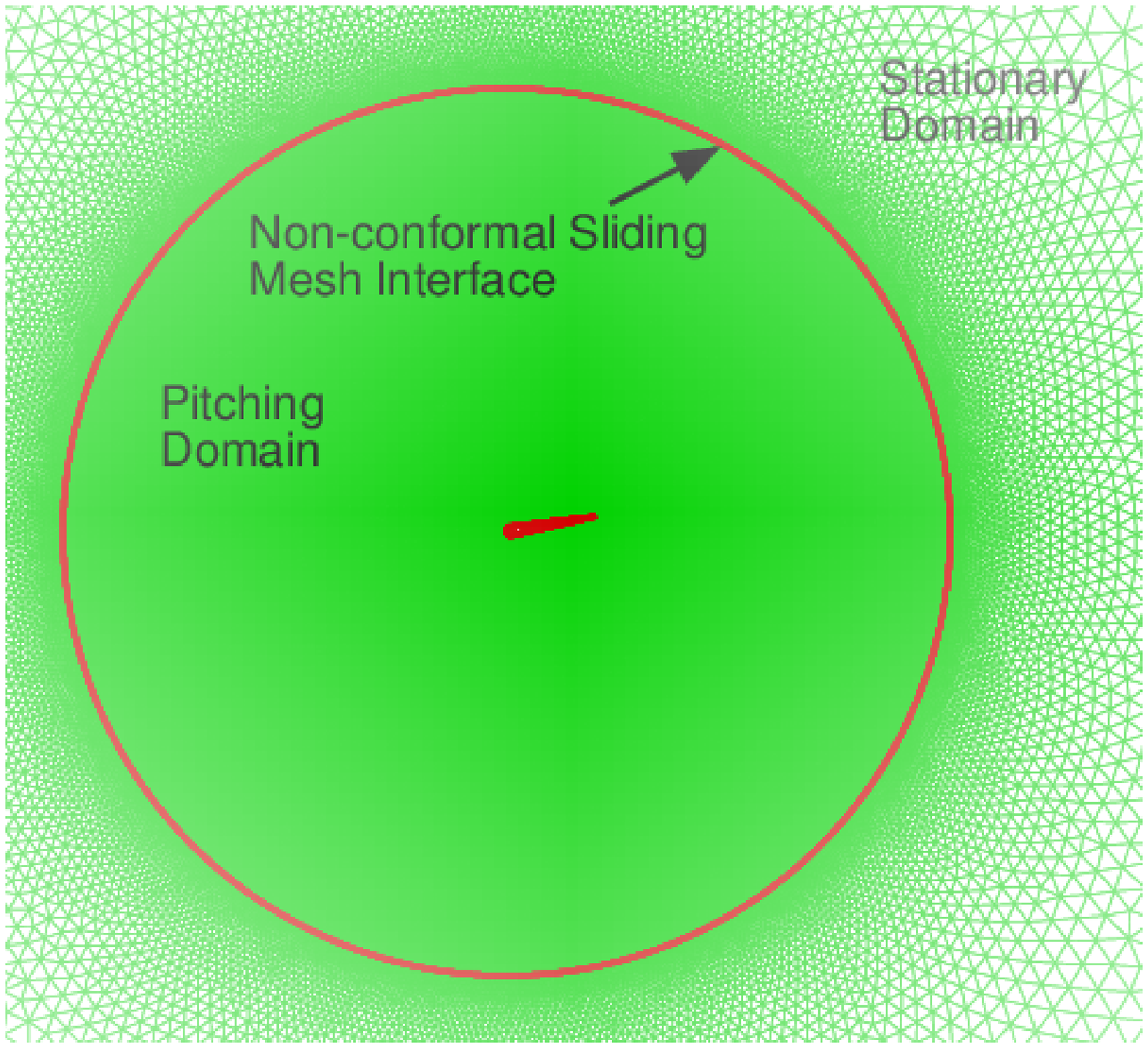}}
\subfigure[]{\label{fig:gridc}\includegraphics[scale=0.2]{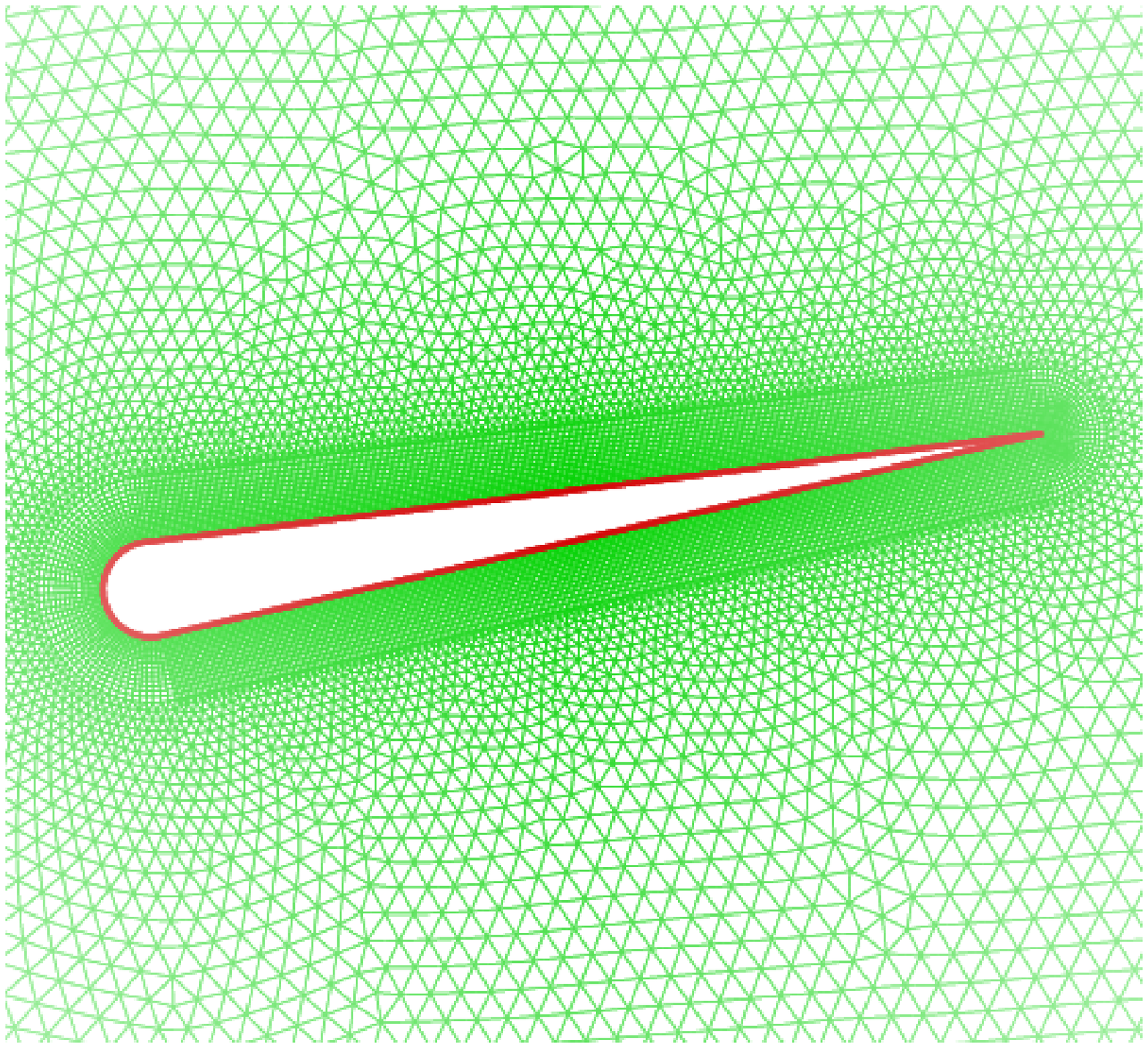}}
\caption{(a) Schematic of computational domain and details of the
boundary conditions, (b) non-conformal sliding interface between the
stationary and the pitching zones, and (c) mesh density around the
whole surface of the airfoil} \label{fig:grid}
\end{figure}

To use higher order schemes for numerical approximation of temporal
term, \cite{Kinsey2008, Kinsey2012a, Kinsey2012b, Kinsey2012c,
Kinsey2014} proposed employment of sliding mesh technique that was
also followed by other researchers like Ashraf et al. \cite{Ashraf2011,
Ashraf2012}. In this technique, pitching airfoil is placed in a
heaving reference frame by applying the time-varying velocity
condition at the inlet boundary, as shown in figure~\ref{fig:grid}. It
serves as the relative velocity of the reference frame with respect
to the airfoil. The pitching motion of the airfoil is attributed to
defining a separate zone around it through non-conformal sliding
mesh circular interface. Although this method is computationally
more expensive for handling moving reference frame in flow-fields,
yet it provides most accurate simulation methodology because its
ensures the maintainability of a higher quality mesh provided to the solver initially.

To avoid effects of the disturbances on the boundaries of the
computational domain, the inlet boundary is placed at a distance of
$10c$ from the leading-edge of the hydrofoil, while the top and
bottom boundaries are $12.5c$ units away from the foil. The outlet
boundary is kept at a distance of $20c$ from the hydrofoil's
trailing-edge. The radius of the inner pitching domain defining the
boundary for the non-conformal sliding interface, is $5c$ as
presented in figure ~\ref{fig:grida}. A zoomed-in view of the pitching
zone is presented in figure~\ref{fig:gridb} . Flow domain is meshed
using unstructured triangular cells. $550$ nodes are present on
hydrofoil surface while maintaining $y^{+}\sim{1}$. To fully capture the features of boundary layer flow, $14$ arrays of the mesh-nodes are placed around the foil as shown in figure~\ref{fig:gridb}.

The convergence criterion for the iterative solution at each
time-step is set to be $1\times10^{-5}$. Although we obtain
convergence with almost $10-15$ iterations at each time-step,
maximum allowable number of iteration are $50$. We run all the
simulations for $10$ oscillations where the steady-state solution is
approached within $2-3$ oscillation cycles, and provide the
statistical quantities based on the data of the last $7$
oscillations.

\subsection{Acoustic Modeling}
\label{subsec:Acoustic} In this paper, we model the acoustics
features using the Ffowcs-Williams and Hawkings (FW-H) method
\cite{Williams1969}. Having its roots in the earlier work by Lighthill \cite{Lighthill1952}, FW-H model predicts the sound
generated by arbitrarily moving rigid bodies. Lighthill's
fundamental analogy is derived from the continuity (\ref{eq:cont}) and momentum equations (\ref{eq:NS}) of fluid
mechanics. FW-H model essentially presents an inhomogeneous wave
equation \cite{Williams1969, Ansys, Mohamed2016, Ghasemian2015} as
given below.

\begin{eqnarray}
\fl \frac{1}{{a^2}_\circ}\frac{\partial^2 p^{'}}{\partial t^2}-{\nabla^2
p^{'}} = {\frac{\partial^2}{\partial{x_i}\partial{x_j}}\{{T_{ij}}{H(f)}\}} 
- \left({\frac{\partial}{\partial{x_i}}\{[{P_{ij}{n_j}}+{{\rho}{u_i}(u_n-v_n)}]\delta(f)\}}\right) \nonumber\\
+  \left({\frac{\partial}{\partial{t}}\{[{\rho_\circ}{v_n}+{\rho}(u_n-v_n)]\delta(f)\}}
\right)
\label{eqn:FWH}
\end{eqnarray}


where $u_i$, $u_n$ are the fluid Cartesian velocity components in
$x_i$ direction, and normal to the integration surface ($f=0$),
respectively. $v_n$ is the velocity of the integration surface
normal to the surface. $\delta(f)$ and $H(f)$ denote the Dirac Delta
and Heavyside functions, respectively. $p$ shows the sound pressure
level in the far-field, and $T_{ij}$ represents the Lighthill's
stress tensor. The unit normal vector $n_i$ points towards the
exterior region ($f>0$), and $a_\circ$ denotes the speed of sound in
the concerned physical medium. The subscript $\circ$ indicates the
free-stream quantities, and the primed quantities shows the
difference between the values in the real domain and the undisturbed
medium \cite{Ghasemian2015}. A mathematical surface ($f=0$) is
introduced to investigate the exterior flow phenomenon ($f>0$) in an
unbounded space. The surface $f=0$ defines the shape and motion of
the source surface which is the hydrofoil in the
present case. $f>0$ corresponds to the interior of the source
surfaces. The terms of the right hand side in \ref{eqn:FWH}
models various source terms for sound generation. The Lighthill
stress tensor, $T_{ij}$, the first source term, is defined as,

\begin{eqnarray}
T_{ij}&={{\rho}{u_i}{u_j}}+{P_{ij}}-{{{a^2}_\circ}{(\rho-{\rho_{\circ}})}{\delta_{ij}}}
\label{eq:Lighthill}
\end{eqnarray}


\noindent where ${P_ij}$ shows the compressive stress tensor; a
combination of the surface pressure ($p$) and the viscous stress.

\begin{eqnarray}
P_{ij}&={p}{\delta_{ij}}-{}
{\mu}\{{\frac{\partial{u_i}}{\partial{x_j}}}+{\frac{\partial{u_j}}{\partial{x_i}}}-{\frac{2}{3}}{\frac{\partial{u_k}}{\partial{x_k}}}\delta_{ij}\}
\label{eq:Pij}
\end{eqnarray}


The source modeled by the Lighthill stress tensor is known as
quadrupole, where $H(f)$ depicts volume distribution of sources in
the outer domain. The second term in \ref{eqn:FWH} models the distribution of dipoles resulting from the unsteady external forces produced due to
the fluid-structure interaction. Last term here describes the
distribution of monopoles generated from displacement of the fluid as a
result of structural motion. The distributions of monopole,
dipole, and quadrupole are also known as thickness noise, loading
noise, and turbulence-induced noise, respectively. Basically, these
are surface distribution sources as identified by the presence of
the function $\delta(f)$ in the relevant terms of \ref{eqn:FWH}.
The inhomogeneous wave equation (\ref{eqn:FWH}) can be integrated
analytically assuming the free-space flow, and the absence of
obstacles between the sound sources and the receivers. The surface
integrals gives contribution from the monopoles and dipoles sources
and partially from quadrupole sources. The volume integrals denote
quadrupole sources in the region outside the source surface. In case
of very low Mach number flows ($M << 1$) such as the present fluid flow phenomenon, volume integrals do not contribute much towards
quantification of noise. Hence, these terms can be neglected in
further computations. Resultantly, we decompose the acoustic pressure
($p^{'}$) as follows.

\begin{eqnarray}
{p^{'}(\vec{x},t)}&={p^{'}_T(\vec{x},t)}+{p^{'}_L(\vec{x},t)}
\label{eq:FWH1}
\end{eqnarray}


\noindent where $\vec{x}$ represents the receiver's position, $t$ is
the observer time, whereas $T$ and $L$ denote the quantities
associated with the thickness noise, and the loading noise,
respectively. Then, we have \cite{Ghasemian2015, Mohamed2014},

\begin{eqnarray}
\fl 4{\pi}{p^{'}_T(\vec{x},t)} = {\int_{f=o}{[\frac{{\rho_{\circ}}({\dot{U}_n}-U_{\dot{n}})}{r{(1-{M_r})}^2}]}_{ret}}dS
- \left({\int_{f=0}{[\frac{{\rho_{\circ}}{U_n}({r}{\dot{M}_r}+{{a_\circ}{(M_r+M^2)}})}{{r^2}{(1-{M_r})}^3}]}_{ret}}dS\right)
\label{eqn:FWH2}
\end{eqnarray}


\noindent and

\begin{eqnarray}
\fl 4{\pi}{p^{'}_L(\vec{x},t)} = \frac{1}{a_\circ}{\int_{f=0}{[\frac{\dot{L}_r}{r{(1-M_r)}^2}]}_{ret}}dS 
+ \left({\int_{f=0}{[\frac{L_r-{L_M}}{r^2{(1-M_r)}^3}]}_{ret}}dS\right) \nonumber \\
- \left(\frac{1}{a_\circ}{\int_{f=0}{[\frac{{L_r}({r}{\dot{M}_r}+{{a_\circ}{(M_r+M^2)}})}{{r^2}{(1-{M_r})}^3}]}_{ret}}dS\right)
\label{eqn:FWH3}
\end{eqnarray}


\noindent where

\begin{eqnarray}
{U_i}&={v_i}+\frac{\rho}{\rho_\circ}(u_i-v_i) 
\label{eq:FWH4}
\end{eqnarray}


\begin{eqnarray}
L_i&=P_{ij}{\hat{n}_j}+{\rho{u_i}}(u_n-v_n) 
\label{eq:FWH5}
\end{eqnarray}


The subscripted quantities in \ref{eqn:FWH2} and \ref{eqn:FWH3}
show the inner products of a vector and a unit vector denoted by the
subscript. For example, $L_r = \vec{L}{\cdot}{\vec{r}}$, and
$U_n={\vec{U}{\cdot}{\vec{n}}}$, where $\vec{r}$, and $\vec{n}$ are
the unit vectors in the radial and normal direction to the wall boundary,
respectively. The dot over a symbol represents the source-time
differentiation of that certain variable. The Mach number ($M_i$)
denotes the ratio of the local surface velocity and the free-stream
speed of sound in the relevant medium. The subscript (${ret}$) shows
that evaluation of the integrand at the retarded time ($\tau$). With $t$
as the observer/receiver time, and $r$ as the distance to the
receiver, it is defined as \cite{Ghasemian2015, Mohamed2014},

\begin{eqnarray}
\tau&=t-\frac{r}{a_\circ} 
\label{eq:FWH6}
\end{eqnarray}

\begin{figure}
\centering
\includegraphics[scale=0.35]{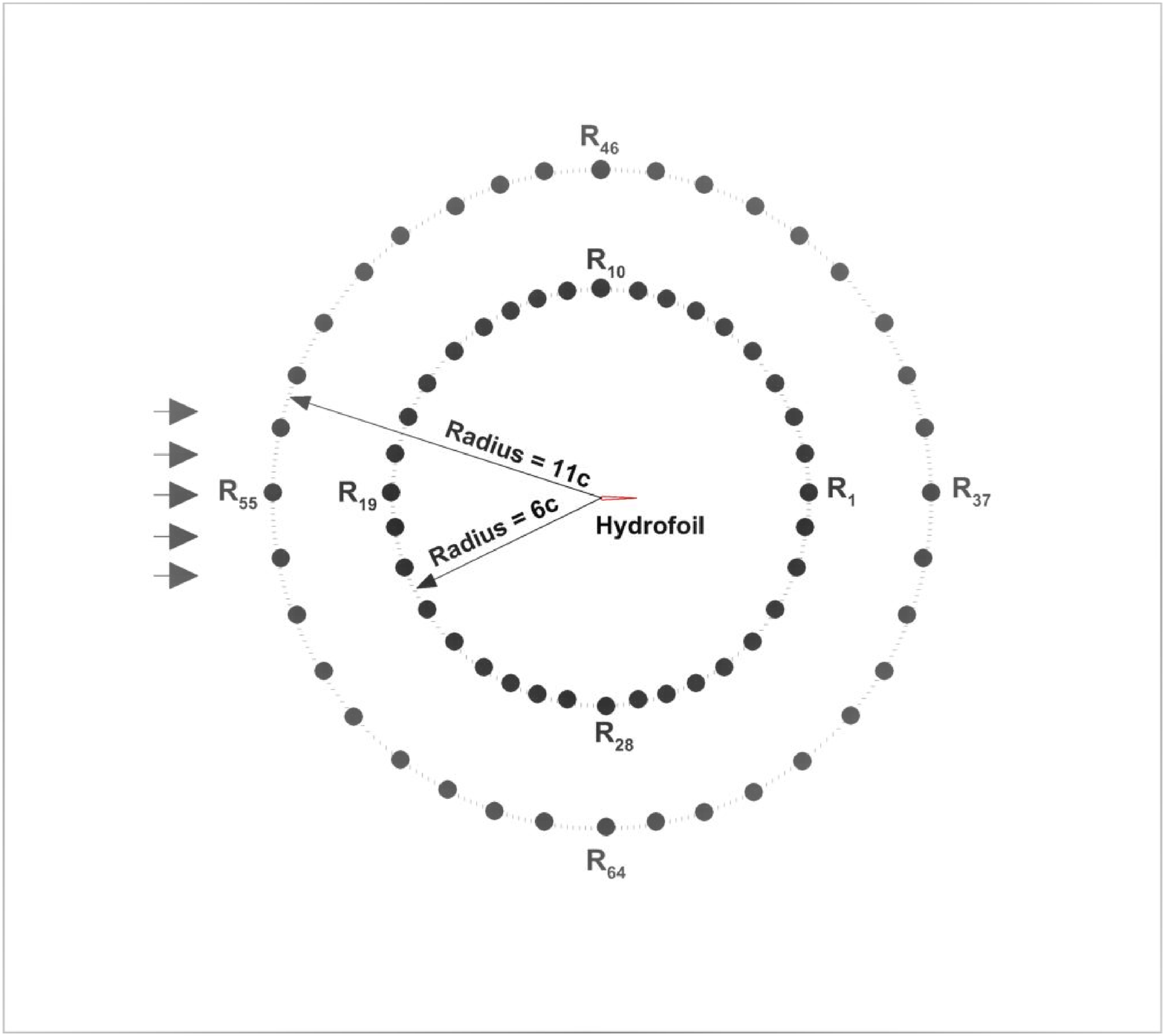}
\caption{Schematic for the placement of acoustic receivers around
the oscillating body} \label{fig:acoustic}
\end{figure}

We record the pressure fluctuations through placement of $72$
acoustic receivers in the surroundings of the hydrofoil as shown in figure
\ref{fig:acoustic}. Due to the spherical propagation (circular in
case of $\mbox{2D}$) of sound from its source, we select two locations to
define circles used for the distribution of of these probes. The distance of these receivers from the pitching-axis of the fish leading-edge are taken as $x/c=6$ and $11$. This positioning of receivers helps determine the directionality in propagation of the sound waves. The receivers are placed at these distances to allow the wake to settle down, and to avoid the recording of the pseudo-sound generated due to the transitional wake very close to the oscillating body. The flow noise caused due to the complex interaction between the near-field flow structures is termed as pseudo-sound \cite{Geng2018}.

\subsection{Performance Parameters}
To measure the hydrodynamic performance of the oscillating hydrofoil, we
compute non-dimensional side-force coefficients, denoted as $C_Y$, $C_D$ and $C_M$ respectively, for all the cases. These coefficient are defined as,

\begin{equation}
\eqalign{C_Y = \frac{F_Y}{{q}{c}} \cr
C_D = \frac{F_D}{{q}{c}} \cr
C_M = \frac{M}{{q}{c^2}}}
\end{equation}


\noindent where $F_Y$, $F_D$, and $M$ denote the side-force, drag. and moment, respectively. $c$ is the chord length of the hydrofoil, and $q={\rho}{{U^2}_\infty}/{2}$ represents the dynamic pressure. Using the time-period ($\tau=1/f_E$) of one pitching cycle, corresponding time-averaged
coefficients are computed using the following relation;

\begin{eqnarray}
\bar{C}&={\frac{1}{\tau}}{\int_t^{t+\tau} {C(t)} dt}
\end{eqnarray}


\subsection{Grid \& Time-step Convergence Study}
Due to the involvement of different Reynolds numbers in our present
study, we perform grid independence study on $\mbox{Re}=5000$ and
$20000$. We compute the hydrodynamic coefficients for the lift ($C_Y$) equal to the side-force presently, thrust ($C_T$), and moment ($C_M$) for three configurations of different mesh sizes. The magnitude of the thrust force is equal to the drag, but these are in opposite directions.
We characterize these configurations as $G_1$ (fine), $G_2$ (medium), and $G_3$ (coarse) as shown in Table~\ref{tab:grid_indp}. It also presents $\bar{C_T}$ for these grids. As the appropriate time-step size is a function of the oscillation frequency in such scenario, we use $2,000$ and $4,000$ time steps in one oscillation cycle to check its convergence. The
results reported in Table~\ref{tab:grid_indp} show reasonable
convergence of $\bar{C_T}$.

\begin{table}
\caption{Results for grid and time-step convergence study at $\mbox{St}=0.50$ and $k=1$}
\centering
\begin{tabular}{c rrrr}
\hline \hline
$\mbox{Re}$ & Grid & No. of Elements & Time-steps per oscillation cycle $\bar{C_T}$ \\
\hline
4700 & Fine ($G_1$) & 393675 & 2000 & 0.7251 \\
4700 & Medium ($G_2$) & 191515 & 2000 & 0.7156  \\
4700 & Coarse ($G_3$) & 94975 & 2000 & 0.7103 \\
\hline
20000 & Fine ($G_1$) & 401110 & 4000 & 0.7251 \\
20000 & Fine ($G_1$) & 401110 & 2000 & 0.7251 \\
20000 & Medium ($G_2$) & 226722 & 2000 & 0.7156  \\
20000 & Coarse ($G_3$) & 99840 & 2000 & 0.7103 \\
\hline
\end{tabular}
\label{tab:grid_indp}
\end{table}

\begin{figure}
\center
\includegraphics[scale=0.6]{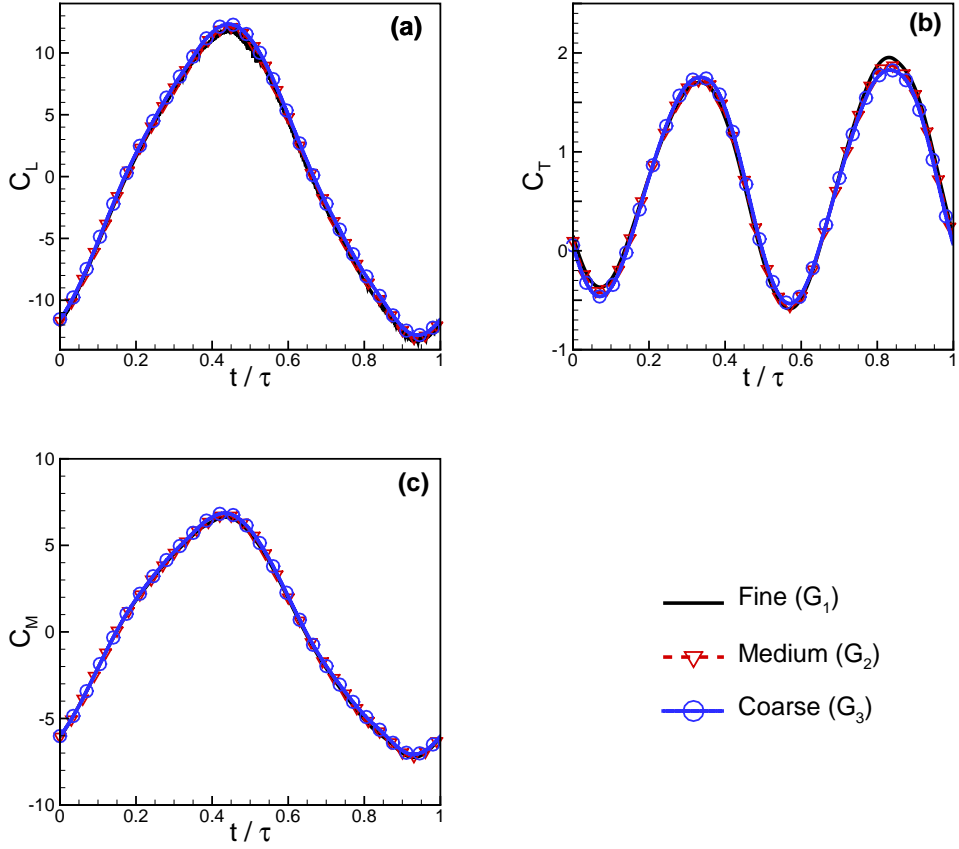}
\caption{Results of the grid independence study for $\mbox{Re}=5,000$}
\label{fig:aero_grid_5000}
\end{figure}

\begin{figure}
\center
\includegraphics[scale=0.6]{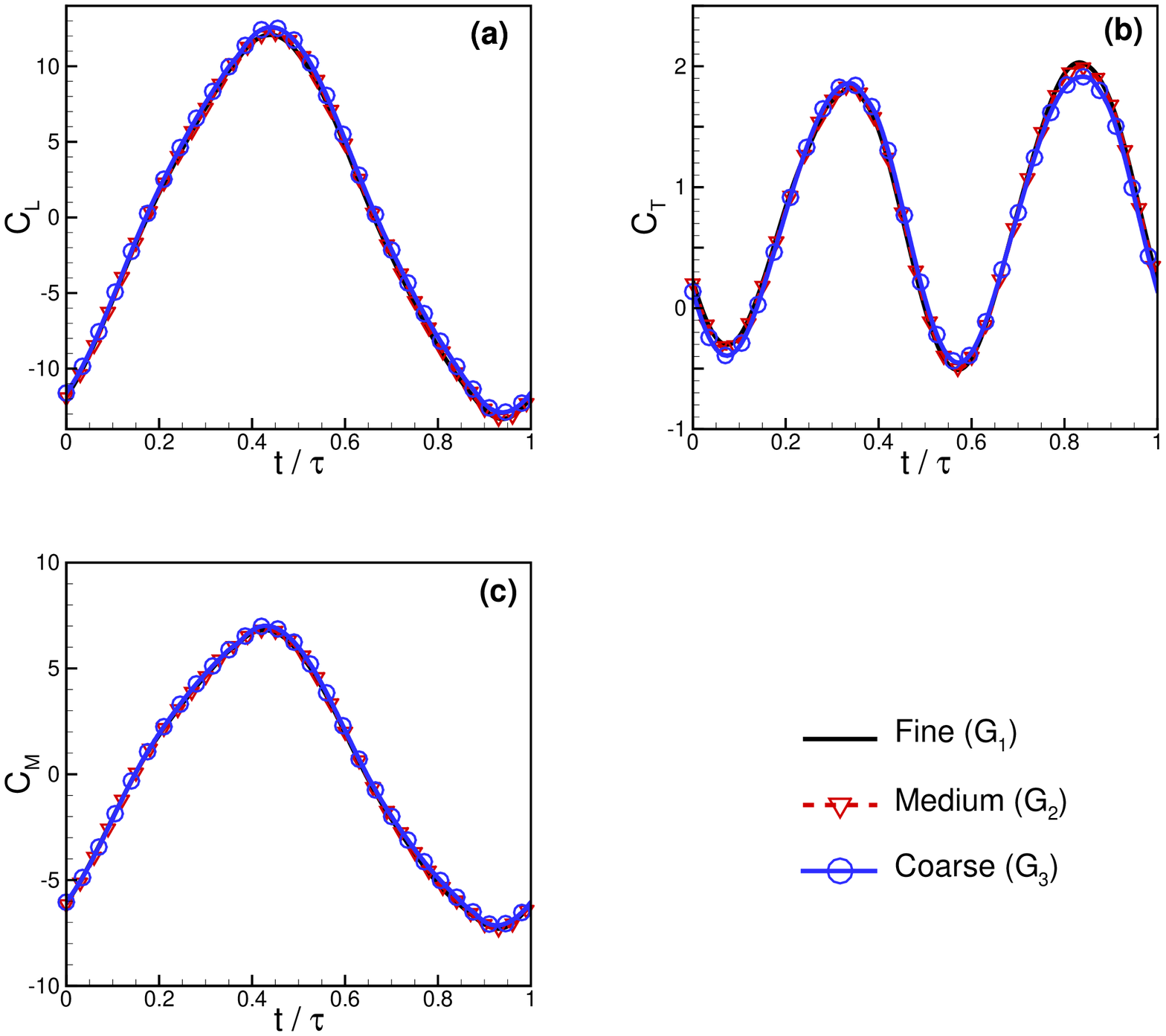}
\caption{Results of the grid independence study for $\mbox{Re}=20,000$} \label{fig:aero_grid_20000}
\end{figure}

\begin{figure}
\center
\subfigure[]{\label{fig:gridc}\includegraphics[scale=0.3]{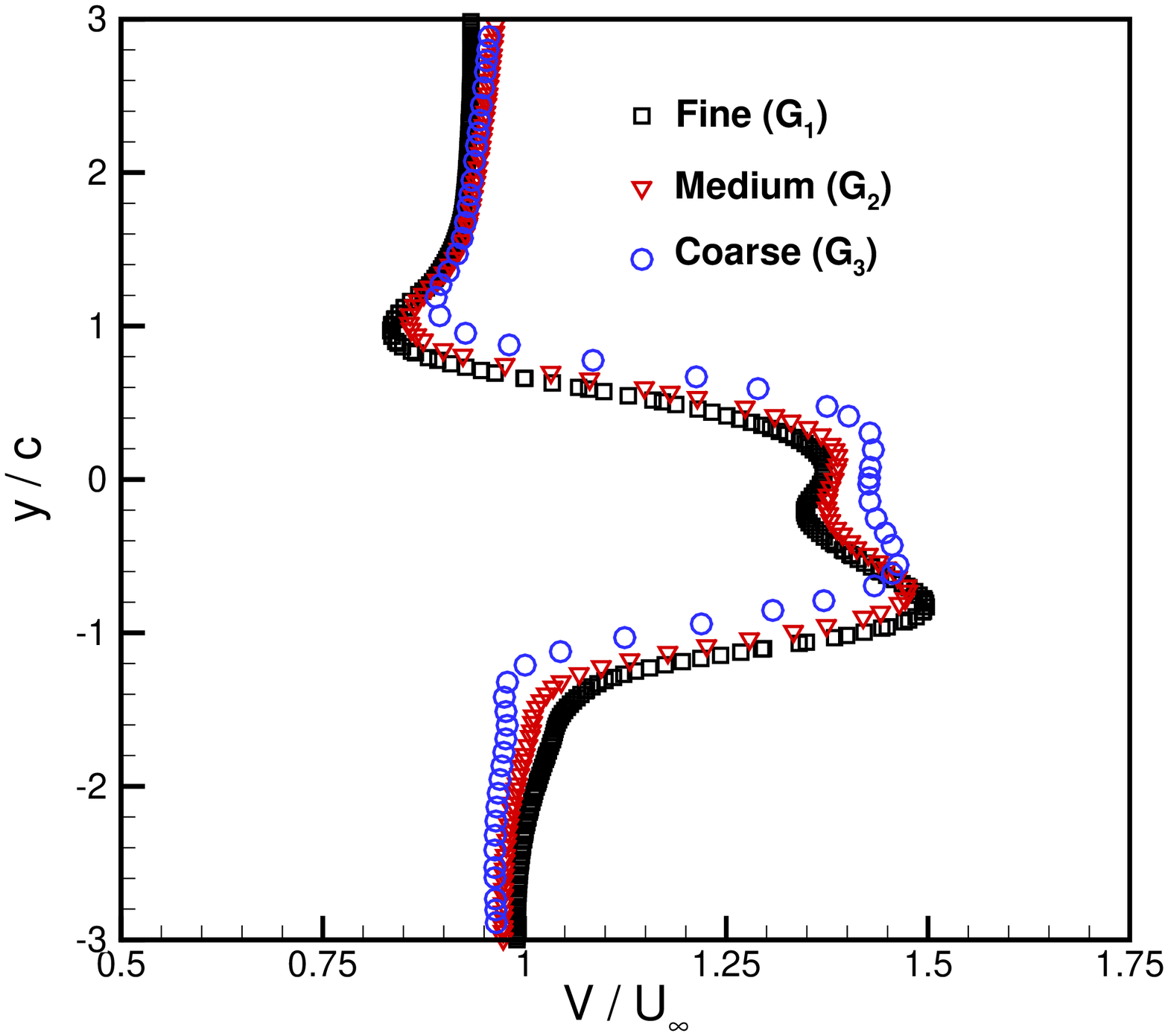}}
\subfigure[]{\label{fig:gridc}\includegraphics[scale=0.3]{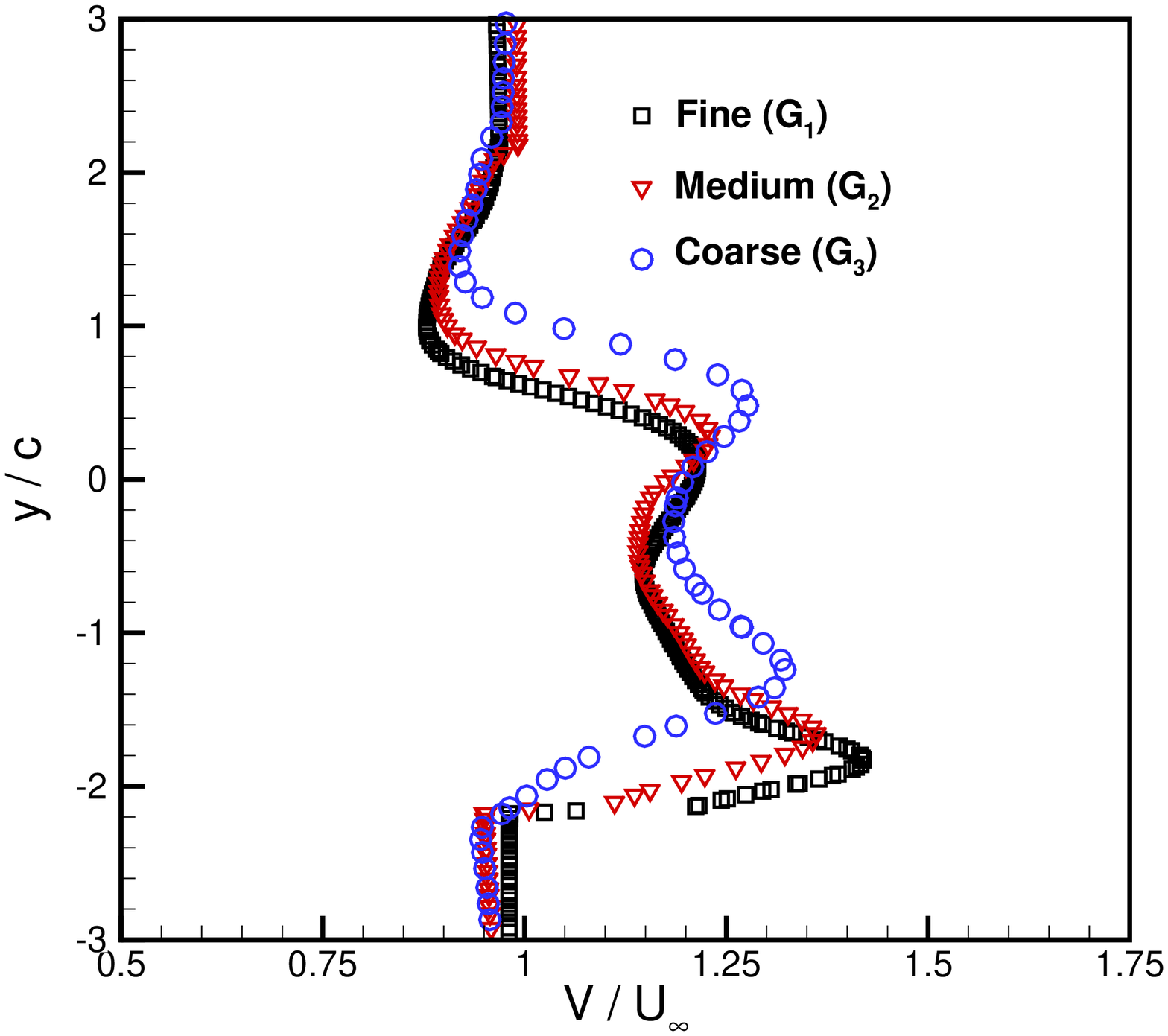}}
\subfigure[]{\label{fig:gridc}\includegraphics[scale=0.3]{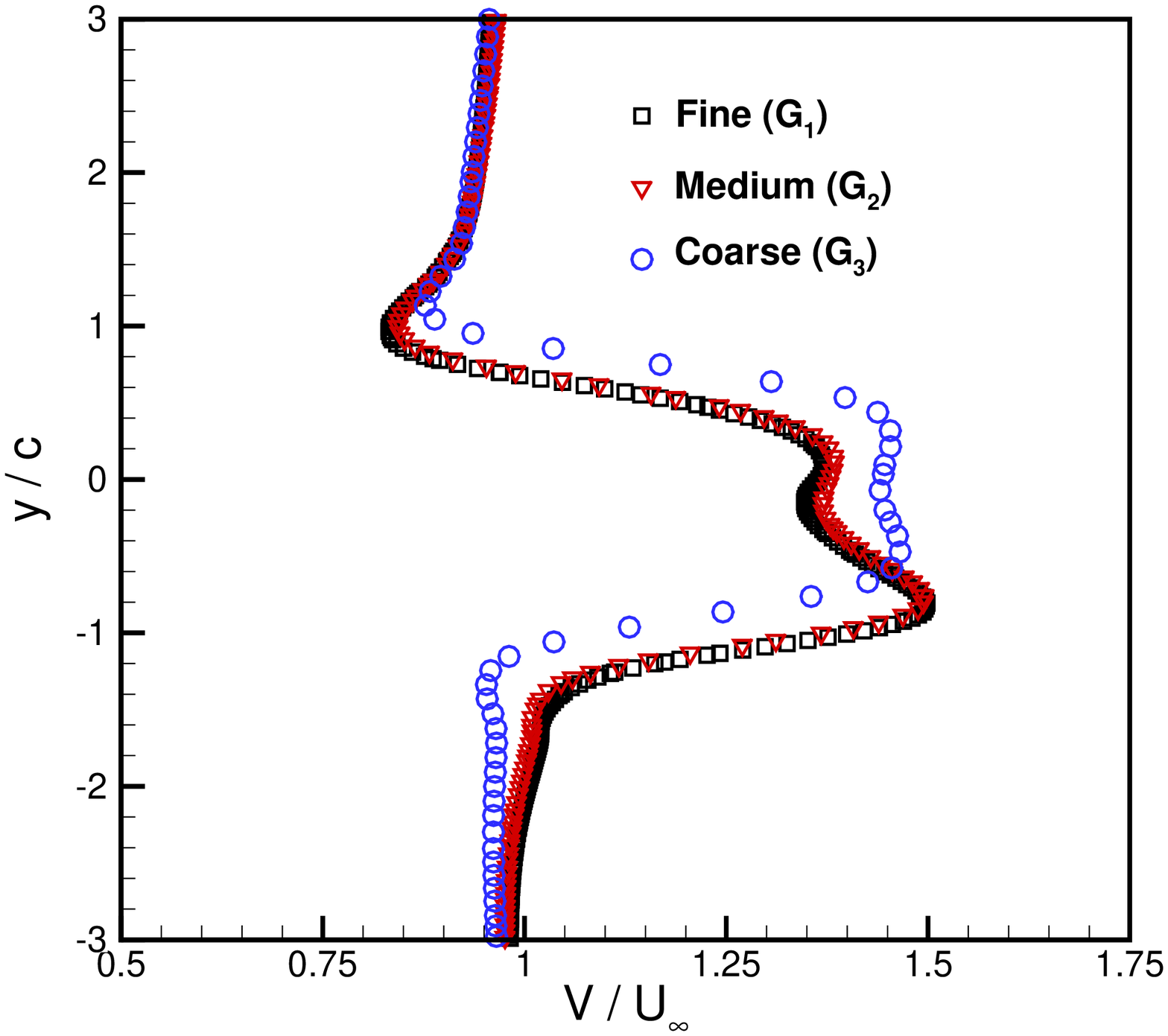}}
\subfigure[]{\label{fig:gridc}\includegraphics[scale=0.3]{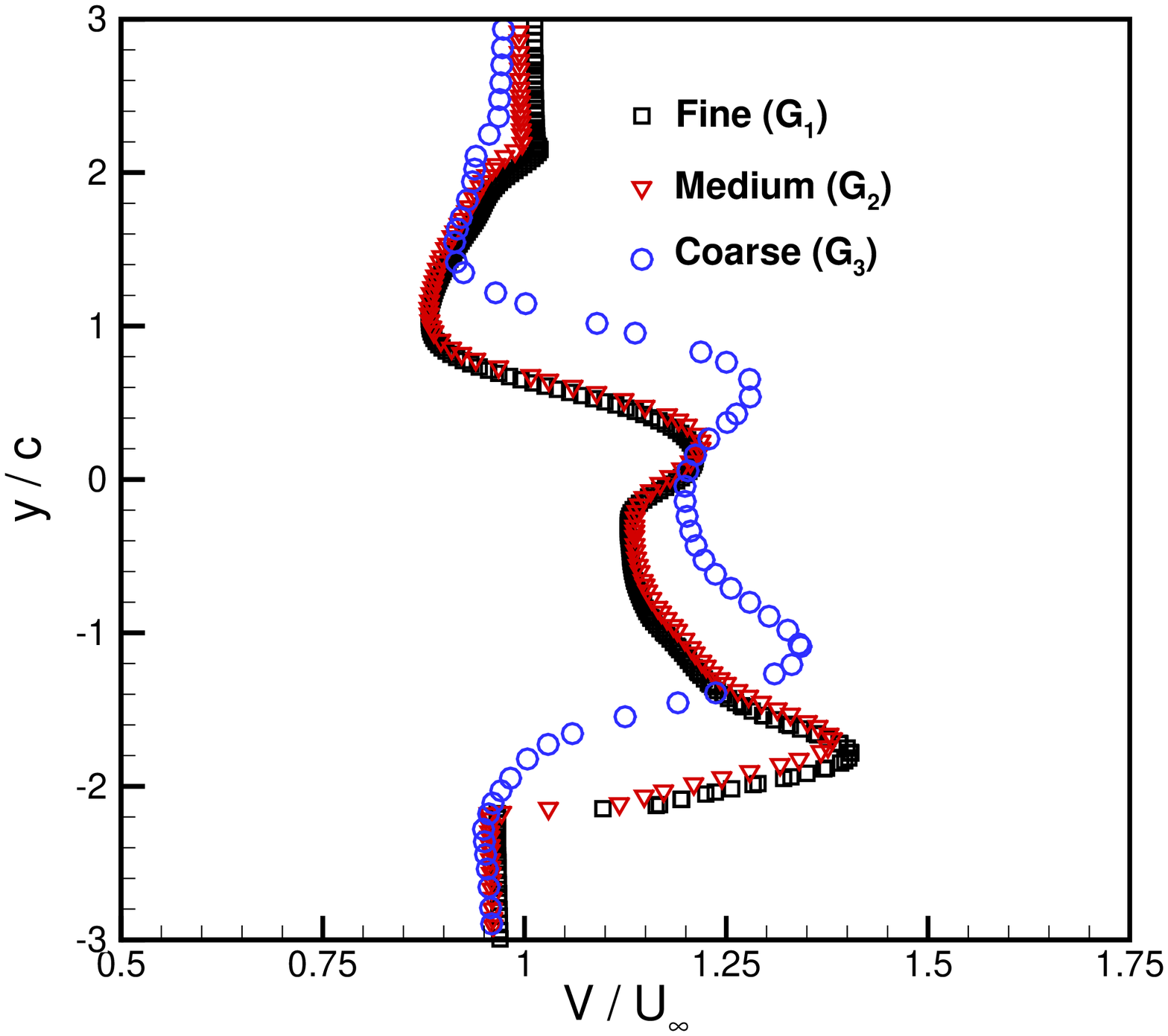}}
\caption{Nondimensional time-averaged velocity profiles in the wake
of oscillating hydrofoil, (a) and (b) are computed at
$\mbox{Re}=5,000$, whereas (c) and (d) show corresponding values for
$\mbox{Re}=20,000$. Left column present values at a distance of
$x/c=2.0$ from the trailing-edge of the hydrofoil, and the right
column here show data for the same at $x/c=3.5$.}
\label{fig:vel_grid}
\end{figure}

Moreover, we also present the temporal profiles of $C_L$, $C_T$, and
$C_M$ in figure~\ref{fig:aero_grid_5000} and \ref{fig:aero_grid_20000} for all the grids and time-step sizes for $\mbox{Re}=5,000$ and $20,000$, respectively. We adjust the grids near the hydrofoil to keep $y^{+}=1$ for different $\mbox{Re}$ so that the boundary layer flow features may be appropriately resolved. We also show the time-averaged velocity magnitudes ($V/U_\infty$) in the wake of oscillating hydrofoil at a distance of $x/c=2$ and $3.5$ in figure~\ref{fig:vel_grid}. These trends show a very small difference in the profiles obtained for $G_1$ and $G_2$. Hence, we employ $G_2$ grid with $2,000$ time-steps per oscillation cycle for our further simulations.

\subsection{Validation}
\begin{figure}
\centering
\includegraphics[scale=0.4]{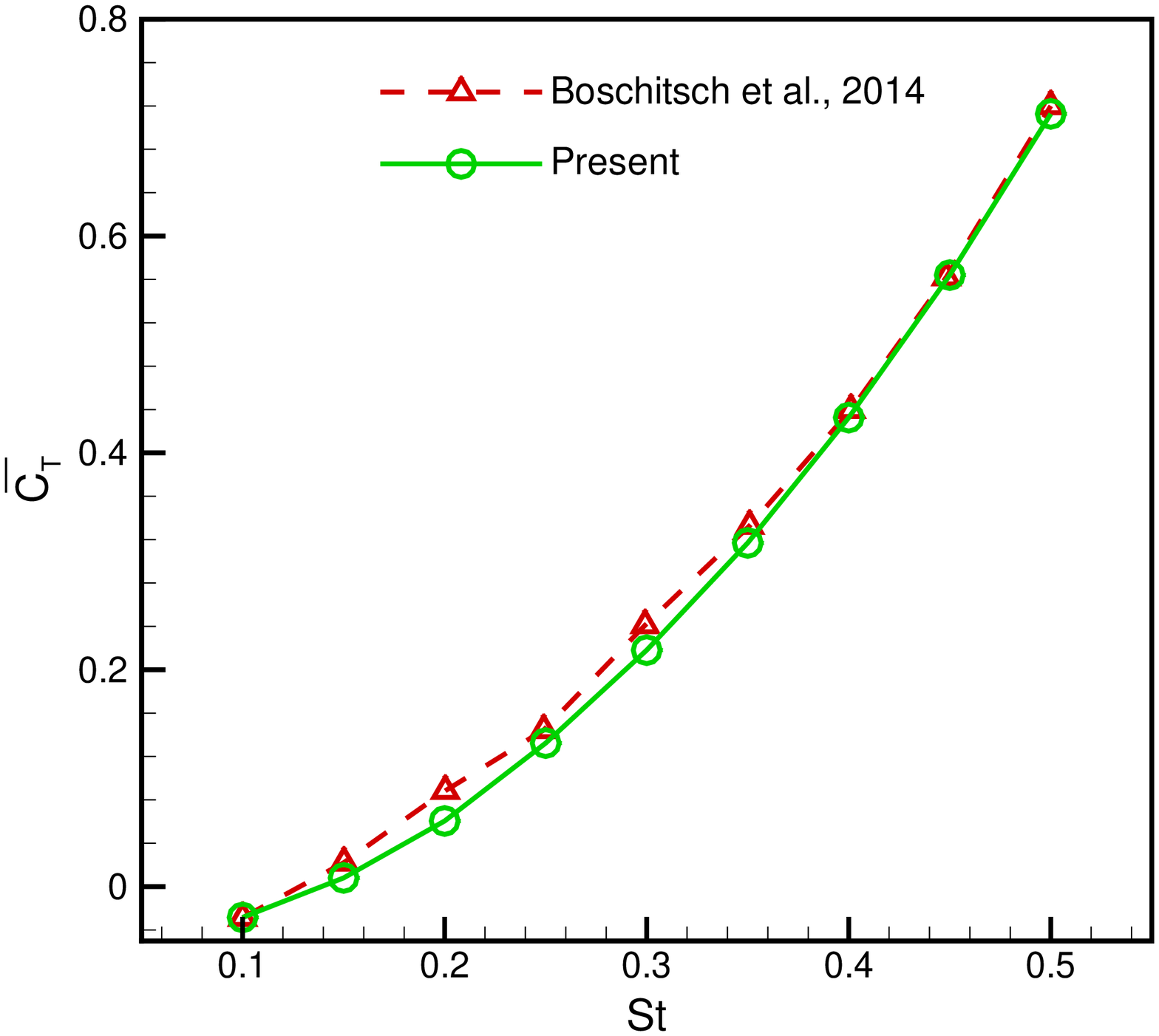}
\caption{Comparison of $\bar{C_T}$ from the present simulations, and
experiments conducted by Boschitsch et al. \cite{Boschitsch2014} in a water tunnel}
\label{fig:valid_2014}
\end{figure}

To validate our current numerical strategy, we compare the
time-averaged thrust coefficients for a range of Strouhal number
with those reported by Boschitsch et al. \cite{Boschitsch2014}. In this
reference study, the same geometrical configuration for hydrofoils
was used in a water tunnel to examine the hydrodynamic performance
of their in-line arrangement. Here, we set $k=1$, $\mbox{Re}={4.7}\times{10^2}$, and $\mbox{St}$ varies from $0.10$ to $0.50$ with an interval of $0.05$. Figure~\ref{fig:valid_2014} shows the comparison between $\bar{C_T}$ for both the studies. It provides a good indication for the accuracy of our simulation methodology.

\section{Results \& Discussion}
In this study, we run all the simulations for 10 oscillation cycles
while utilizing the data for the last $7$ oscillation cycles for the
computation of acoustic features after achieving the steady-state in
the solution of the governing equations. We record the acoustic
pressure signals at the designated receivers/probes, and process them with the Fourier transform technique (FFT) to extract the
frequencies and their contribution towards the generation of these
signals. Next, we use the reference pressure ($p_{ref}=1\times10^{-6}\mbox{Pa}$) for the conversion of the
amplitude of all these frequency components into the sound pressure
levels, commonly known as $\mbox{SPL}$. The reference pressure
usually indicates the minimum sound pressure a human ear can sense
at a certain frequency in water. Here, the role of $\mbox{Re}$, $\mbox{St}$, and $k$ is quantified for the radiation of flow noise and its directivity caused by the oscillating fish-like hydrofoil.

\subsection{Emission of Sound Radiations}
\begin{figure}
\centering
\includegraphics[scale=0.55]{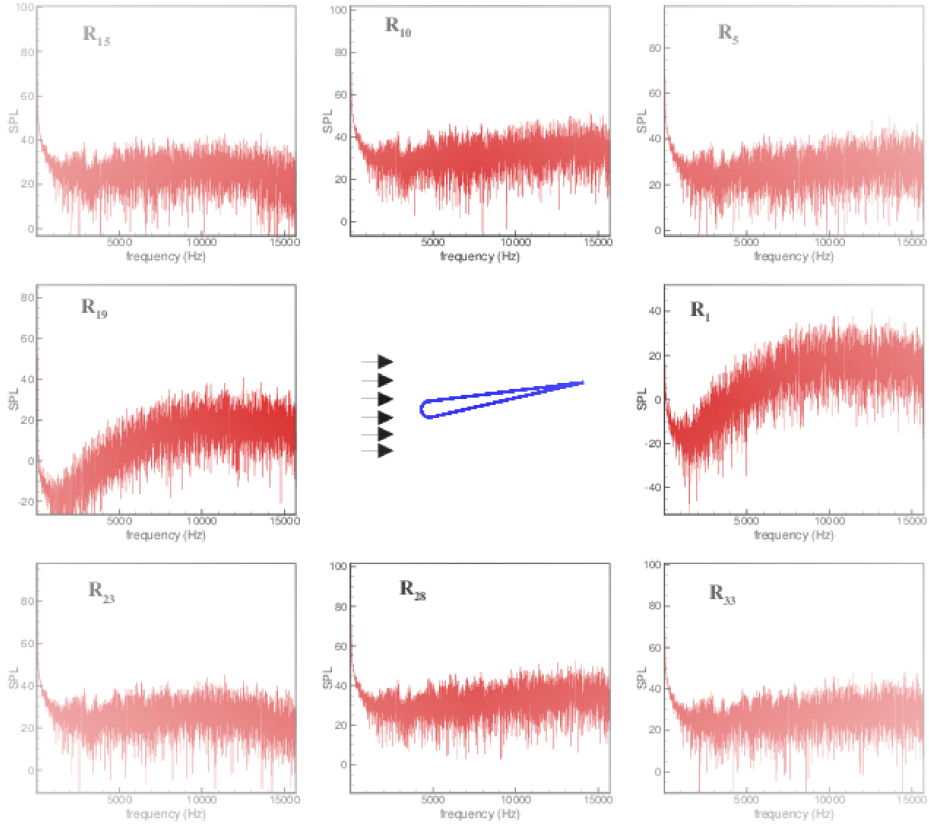}
\caption{Frequency composition of SPL around the pitching hydrofoil
at $8$ locations in the circle closest to the body}
\label{fig:SPL_FFT_1}
\end{figure}

We perform our current numerical simulations for $\mbox{Re}=5,000$,
$10,000$, and $20,000$ while varying $k$ with values of $1$, $2.5$,
and $5$. These values represent lower, middle and higher range of
their associated parameters in hydrodynamics of aquatic species. We
also set $\mbox{St}$ ranging from $0.1$ to $0.50$ which are usually
employed by natural species for swimming to achieve more efficient
performance \cite{Saadat2017}. Figure~\ref{fig:SPL_FFT_1} shows a
representative schematic at $\mbox{Re}=20,000$, $k=5$, and $\mbox{St}=0.50$ for computation of $\mbox{SPL}$ at the probes of the inner circle around the pitching body. Fourier spectra for SPL comes out to be broad-band for the higher range of frequencies in all the cases. It
is also interesting to observe that the forcing frequency has the
most dominant component even in the upstream direction of the flow.
Additionally, these spectra reveal the propagation of pressure
disturbance in all the directions, just like observed previously for
the other flying species \cite{Bae2008, Inada2009, Geng2018}.
Depending on the sensitivity of the hearing parts of the surrounding
bodies, the presence of a fish may be potentially detected by them. Additionally, it is only along the horizontal direction where the strength of the broadband noise in the spectra appears to be almost of the order of that of the forcing frequency and its first even harmonic.

\begin{figure}
\centering
\includegraphics[scale=0.55]{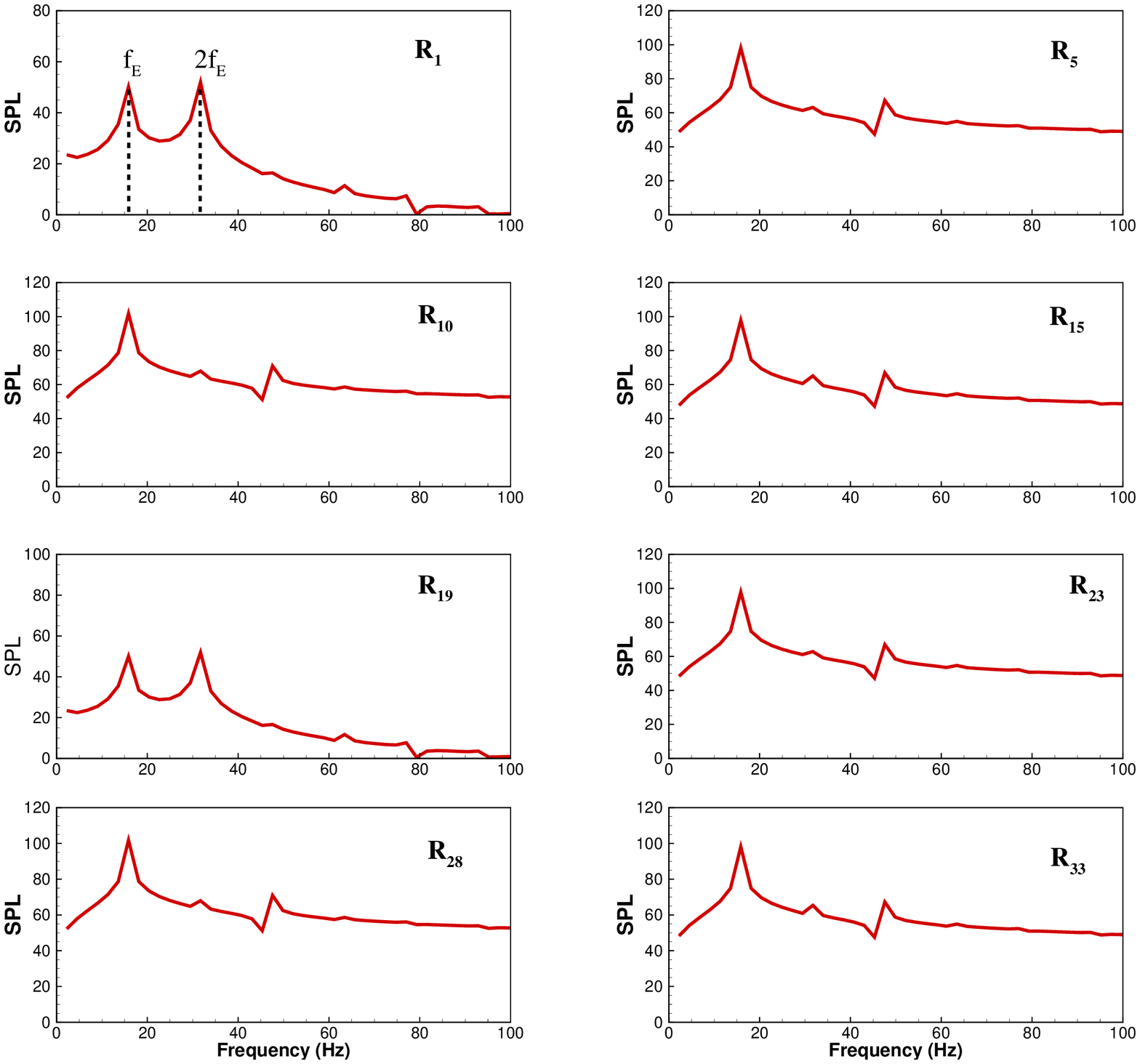}
\caption{Frequency composition of SPL around the pitching hydrofoil
at $8$ locations in the circle closest to the body}
\label{fig:SPL_FFT_2}
\end{figure}

To uncover the spectral features in more detail, we show a zoomed-in
view of these plots in figure~\ref{fig:SPL_FFT_2}. It is obvious that the forcing frequency ($f_E$) and its first even harmonics at $2f_E$
play the strongest role in emission of the flow noise. Their
significance also lies in the fact that these two frequency
components are the strongest in the constitution of the hydrodynamic
side-force and the drag/thrust force, respectively, to be discussed
later in section~\ref{sec:CF}. This trend is evidently seen for all the
receivers. For the sake of brevity, we do not show the higher range
of frequencies that constitutes the broadband noise in the latter
part of these spectra. Some of the higher frequencies have SPL
levels below zero which shows that the associated pressure is less
than the reference sound pressure potentially causing difficulty in
sensing such frequencies. These observations are valid for the
whole range of governing kinematic and flow parameters used for our
present investigations.

\begin{figure}
\centering
\includegraphics[scale=0.7]{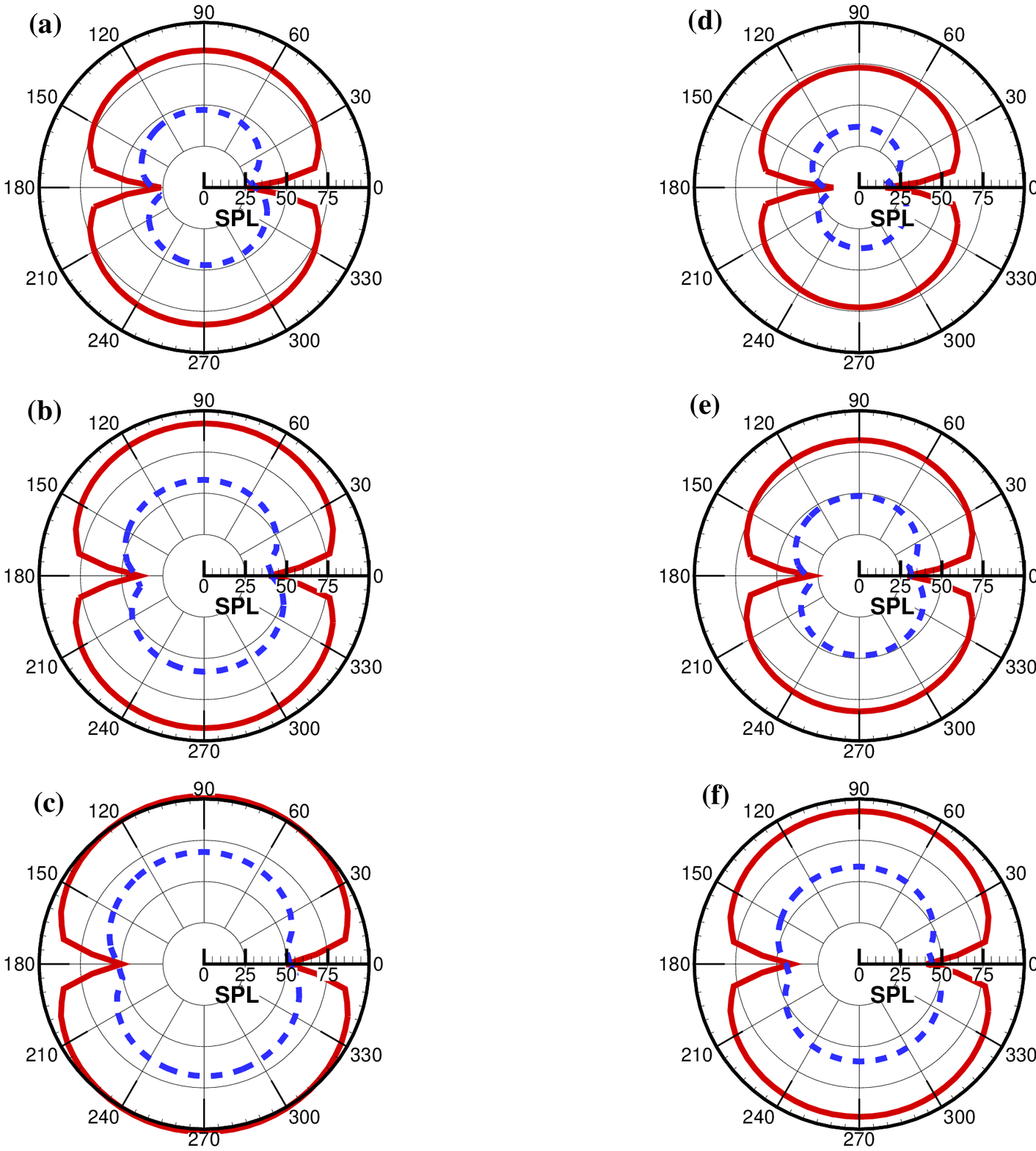}
\caption{Polar plots for distribution of SPL for $k=5$ and
$\mbox{St}=0.50$ where the top, middle and bottom rows show data for
$\mbox{Re}=5,000$, $10,000$ and $20,000$, respectively. Left and
right columns present measurements for SPL at the inner and outer
circles, respectively.} \label{fig:SPL_Polar_1}
\end{figure}

We show a comparison for the measurements in SPL for receivers
placed at the inner and outer circles at a distance of $6c$ and
$11c$ from the pitching-axis locations, respectively, in
Figure~\ref{fig:SPL_Polar_1}. These sound signals tend to dissipate quickly as they traverse in the surroundings. The parameters, $k=5$ and $\mbox{St}=0.50$, here represent the strongest disturbance in the flow media both in terms of oscillation amplitude and frequency. All the
polar plots are drawn on the same scale for ease in comparison. With
increasing $\mbox{Re}$, the SPL grows for both the frequencies,
$f_E$ and $2f_E$. We also notice that the propagation of sound
radiations depicts the pitching body as a dipolar sound source for
both the most dominant frequencies. As we move away from the body,
the magnitudes of SPL get reduced for both the frequencies. Relying on
such findings, our further discussion is based on the results
computed at the receivers placed on the inner circle. Regarding the
pattern of sound propagation, it is a dipole-like
structure for $f_E$ and $2f_E$. Although the peak-valley difference
in this pattern for $2f_E$ is smaller as compared to
that for $f_E$, yet the axis direction is clearly identified. Our
observations for the SPL patterns are in agreement with those of \cite{Geng2018}. It is revealed that the $f_E$ dominates the
sound spectra in all the directions, except along the horizontal
axis. This trend persists for all the parameters considered
presently.

\begin{figure}
\centering
\includegraphics[scale=0.7]{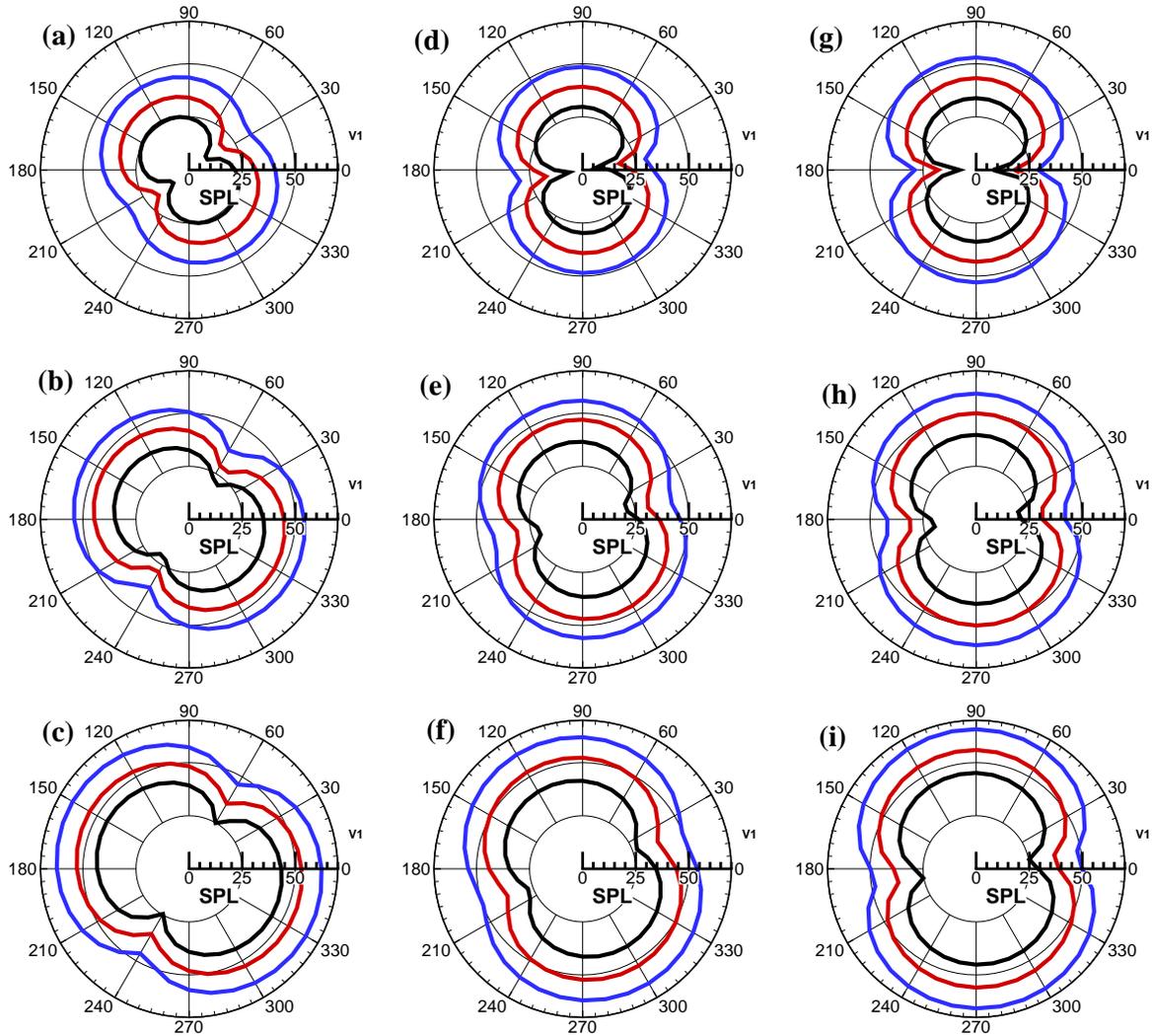}
\caption{Polar plots for distribution of SPL for $k=1$, $2.5$ and
$5$ whereas $\mbox{St}=0.15$, $0.30$ and $0.45$. Here the top,
middle and bottom rows show data for $\mbox{St}=0.15$, $0.30$ and
$0.45$, respectively. Left, middle and right columns present
measurements for SPL at $k=1$, $2.5$ and $5$, respectively. Aiming
for clarity, we prepare these plots using solid lines, and the
readers are referred to the online version for colored figures. In
each of these plots, black, blue and red lines show the SPL patterns
for $\mbox{Re}=5,000$, $10,000$ and $20,000$, respectively.}
\label{fig:SPL_Polar_2}
\end{figure}

Moreover, we find that the dipole axis for SPL at $f_E$ remains
horizontal for the whole range of $\mbox{Re}$, $k$, and $\mbox{St}$. None
of these parameters affect the direction of this dipole axis.
However, the dipole axis for SPL at $2f_E$ is sensitive to these
parameters, and changes its angular position. To inspect it further,
we extract and present the polar plots for SPL distributions for a
range of all the governing parameters in figure~\ref{fig:SPL_Polar_2}.
The SPL radiation pattern in figure~\ref{fig:SPL_Polar_2}a for $k=1$,
$\mbox{St}=0.15$ and $\mbox{Re}=20,000$, and in figure~\ref{fig:SPL_Polar_2}f for $k=2.5$, $\mbox{St}=0.45$, and $\mbox{Re}=20,000$ possess a negligible peak-valley difference, but a slight deviation in these dictating parameters provides evidence that these are essentially a representation of dipole-like structures. Contrary to our
observations here, Geng et al. \cite{Geng2018} opined that these
kind of pattern looked like a monopole. Here, we support our
argument by the parametric study revealing the dipolar nature of the SPL
radiations at $2f_E$ in case of oscillating bodies for a wide range
of governing $k$, $\mbox{St}$, and $\mbox{Re}$. It may need more
investigations whether even larger $\mbox{Re}$ may convert these
dipole-like SPL distributions into monopole-like ones. Furthermore,
these plots also manifest that increase in $\mbox{Re}$ enhances the
strength of $\mbox{SPL}$ for all $k$ and $\mbox{St}$. Higher values of
$\mbox{St}$ by increasing oscillation amplitude of the hydrofoil
causes growth in magnitudes of the sound pressure levels without
appreciably affecting the angular position of the dipole-axis. For
some constant $\mbox{Re}$ and $\mbox{St}$, variations in $k$ has a
negligible affect on the sound pressure levels, but it emerges as a
significant factor for setting the direction of the dipole-axis.
While looking at the plots in a row in figure~\ref{fig:SPL_Polar_2},
it becomes apparent that an increase in $k$ revolves the dipole-axis
clock-wise.

\subsection{Role of Hydrodynamic Pressure}
\label{sec:CF} 
\begin{figure}
\centering
\includegraphics[scale=0.6]{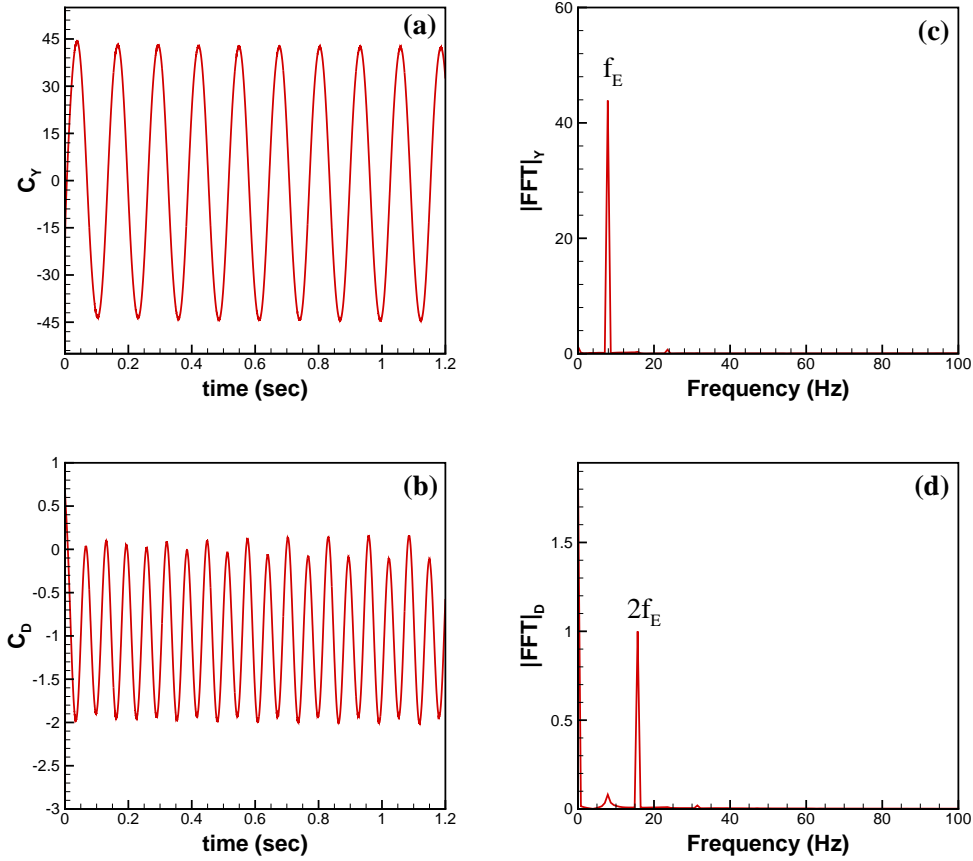}
\caption{Temporal plots of $C_Y$ and $C_D$ are shown in (a) and (b), respectively, for $\mbox{Re}=10,000$, $k=5$ and $\mbox{St}=0.50$. (c) and (d) present the Fourier transforms of both the hydrodynamic force coefficients.}
\label{fig:CL_CD_FFT}
\end{figure}

It is a known fact that sound is propagated through
the pressure fluctuations in the surroundings of the noise source.
When a solid body oscillates, it causes fluctuation of pressure on
its both sides from positive to negative, and vice versa. During an
upstroke, the upper surface of the hydrodynamic body appears as the
loading side, and experiences positive pressure while displacing the
water around. In this case, negative pressure is seen at the lower
surface. On the other hand, the signs of the pressure signals get
reversed during a down-stroke. These oscillatory pressure quantities
are the responsible factors for the flow noise to emit from the body
and radiate in the surroundings. Previously, some investigations \cite{Sueur2005, Bae2008, Inada2009, Lu2014, Geng2018} focused towards establishing connections of the SPL distribution patterns with the fluid force coefficients. These coefficients, $C_Y$ and $C_D$, are presented as
the global characteristics to measure pressure perturbations, the loadings on the body, and its hydrodynamic performance.
Similarly, we also  explore the role of these force coefficients and
their associated frequency compositions in the patterns of sound
pressure levels. Figure~\ref{fig:CL_CD_FFT} displays the temporal
histories and the amplitude spectra for $C_Y$ and $C_D$. It can be
observed that the most dominant frequency in $C_Y$-spectrum is
$f_E$, whereas it is $2f_E$ in $C_D$-spectrum. This feature of fish
swimming has also been revealed earlier by Khalid et al. \cite{Khalid2016a}. This characteristic clearly explains the dominating effect of $2f_E$ in the sound pressure levels along the horizontal direction for the whole range of parameters considered in this study. This observation is also a depiction of the association between the SPL distributions and the hydrodynamic force coefficients.

\begin{figure}
\centering
\includegraphics[scale=0.7]{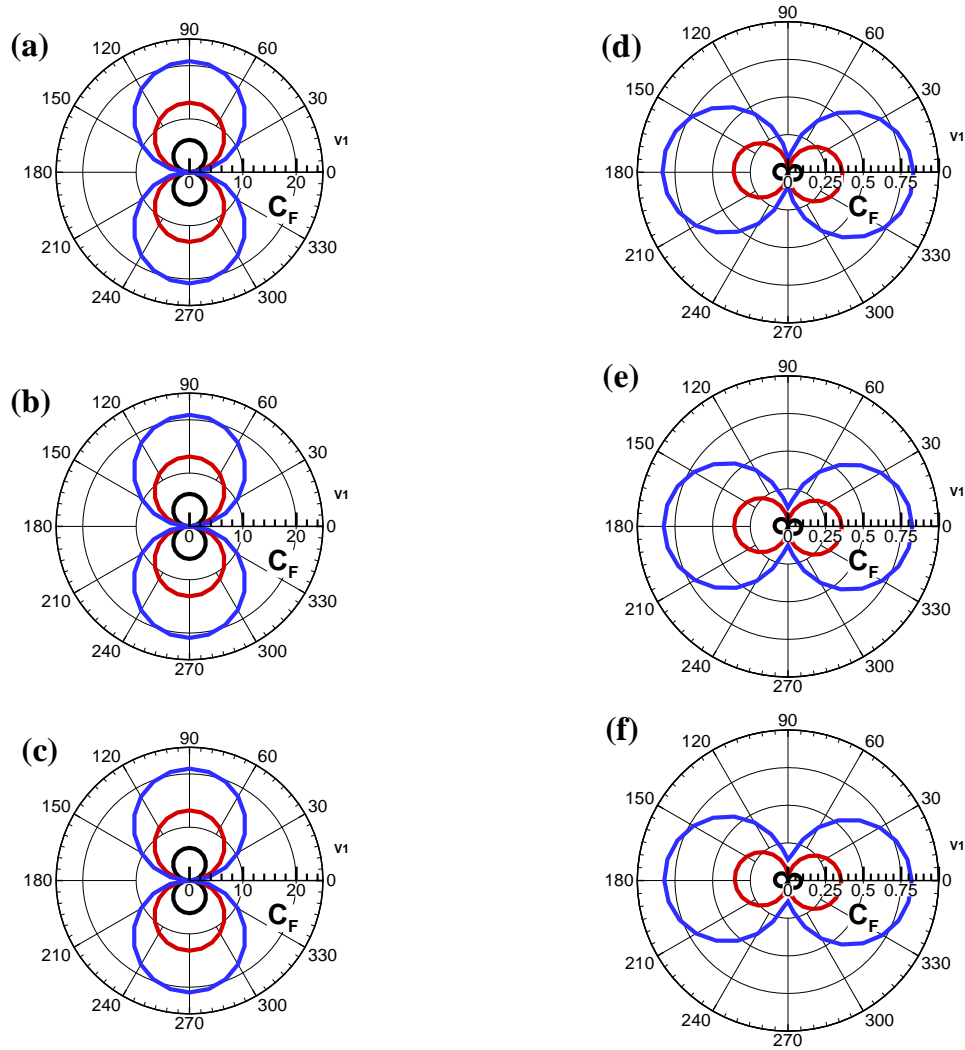}
\caption{Distribution of total hydrodynamic force coefficients for a
range of $\mbox{Re}$, $k$ and $\mbox{St}$. Top, middle, and the
bottom rows show data for $\mbox{Re}=5,000$, $10,000$, and $20,000$,
respectively. Left and right columns belong to $C_F$ distributions
at $f_E$ and $2f_E$, respectively. Black, blue and red lines in each
of the figures show patterns for $\mbox{St}=0.15$, $0.30$ and
$0.45$, respectively.} \label{fig:CF_Plot_1}
\end{figure}

Next, we aim to analyze the directionality and distribution of SPL
in correlation with those of the $C_Y$ and $C_D$. Following the
approach introduced by \cite{Geng2018}, we project the force
coefficient vectors onto the normal directions at each receiver
point on the inner circle using the following relation.

\begin{eqnarray}
C_F(\vec{r},t)&=\{C_D(t),C(y)t\}.\frac{\vec{r}}{|\vec{r}|} 
\label{eq:CF}
\end{eqnarray}


\noindent where $C_F$ denotes the hydrodynamic coefficient in the
normal direction of an acoustic receiver represented by $\vec{r}$
that initiates from the pitching-axis location. Taking the Fast
Fourier transform of $C_F$ and computation of their absolute
amplitudes for the frequencies, $f_E$ and $2f_E$, provide us with the
patterns shown in figure~\ref{fig:CF_Plot_1}. These polar plots
indicate the dipole-like distribution patterns of the overall
hydrodynamic force coefficients for $f_E$ and $2f_E$, and these remain the same for the whole range of parameters considered in this piece
of work. It is also interesting to highlight that the dipole-axis for
$C_F$ patterns at $f_E$ and $2f_E$ are perpendicular to each other.
Here, the dipole-like structures in the left column
(figure~\ref{fig:CF_Plot_1}a-c) showing $C_F$ at $f_E$ matches with
their corresponding SPL distribution patterns, clearly revealing
that both the sound radiation patterns at $f_E$ and their axis
correlate with the force coefficients. However, the plots in the
right column (figure~\ref{fig:CF_Plot_1}d-f) pose similarity in the
distribution patterns with those of the SPL, but the directions of
the dipole-axis do not match. Moreover, we also observe that the
magnitudes of the $C_F$ patterns stays insensitive to the variations
in $\mbox{Re}$, but SPL dipole-like patterns grow at larger
$\mbox{Re}$. Hence, there exist some additional unknown
characteristics responsible for controlling not only the magnitudes
of the sound pressure levels, but also their dipole-axes. We intend
to explore it further in our future work.

\section{Conclusions}
In this study, we perform numerical simulations by modeling a moving
fish body using a pitching hydrofoil for a range of flow and
kinematic parameters. We focus on how these factors affect the
radiation of sound in the surrounding water. We seek to find the
dependence of the magnitude of the sound pressure levels and the
directionality of sound waves through the associated dipole-axes on
these governing parameters. Our parametric investigations lead us to
the following conclusions. For the whole range of Reynolds number,
Strouhal number, and reduced frequency considered in the present
study, distributions of sound pressure levels depicting the flow
noise around the swimming body present dipole-like radiations
patterns for both the oscillation frequency, $f_E$, and its first
even harmonic, $2f_E$. The oscillation frequency dominates in the
whole spectrum as compared to $2f_E$ except along the horizontal
direction. It is in agreement with the directional strengths of the
hydrodynamic forces along the horizontal and sideway directions. As
$C_D$ represent the hydrodynamic force in the horizontal direction,
and the strongest frequency in its spectra is $2f_E$, we see its
dominance in its relevant direction only. Reynolds and Strouhal
numbers affect the magnitude of the sound pressure levels. Both of
these factors do not seem to control the dipole-axis, and their
variations do not show a significant impact in setting its angular
position. On the other hand, reduced frequency appears to be the
controlling feature for the direction of the dipole-axis. It shows
only a minor impact on the magnitude of the SPL levels. Out of all
the considered kinematic and flow parameters, Reynolds number comes
out to be a potential factor to convert the dipole-like sound
radiation pattern into monopole-like ones. There exists a
possibility to witness this phenomenon with even higher values of
$\mbox{Re}$ that required further inspection. Finally, our current
work shows that the fishes and other aquatic animals may control
their acoustic signals to their target bodies through adopting
certain kinematic features.

\section*{Acknowledgments}
Dr. Jiang acknowledges the support from the Independent Innovation
Fund Project of Agricultural Science and Technology in Jiangsu
Province through the Grant No. CX(16)1004, and Key Research and
Development Plan of Jiangsu Province (Modern Agriculture) through
the Grant No. BE2017334.

\section*{References}

\bibliographystyle{iopart-num}
\bibliography{mybibfile}

\end{document}